\documentclass[letterpaper,aps,prc,superscriptaddress,nofootinbib,showpacs,floatfix,reprint]{revtex4-1}
\usepackage[utf8]{inputenc}
\usepackage{amsmath}
\usepackage{amsfonts}
\usepackage{amssymb}
\usepackage{amsthm}
\usepackage{bm}
\usepackage[english]{babel}
\usepackage{fontenc}
\usepackage{graphicx}
\usepackage[margin=1in]{geometry}
\usepackage{enumerate}
\usepackage{color}

\begin{document}

\title{Quark Flavor Balancing in Nuclear Collisions}
\author{Yash Patley}
\email{yashpatley.iitb@gmail.com}
\author{Basanta Nandi}
\email{basanta@phy.iitb.ac.in}
\author{ Sadhana Dash }
\email{sadhana@phy.iitb.ac.in}
\affiliation{Department of Physics, Indian Institute of Technology
 Bombay, Mumbai - 400076, India }
\author{ Victor~Gonzalez }
\email{victor.gonzalez@cern.ch}
\affiliation{Department of Physics and Astronomy, Wayne State University, Detroit, 48201, USA}
\author{ Claude Pruneau }
\email{claude.pruneau@wayne.edu}
\affiliation{Department of Physics and Astronomy, Wayne State University, Detroit, 48201, USA}

\date{\today}

\begin{abstract}
The notion of charge balance function, originally designed to study the evolution of charge production in heavy-ion collisions, is extended to consider quark flavor balancing. This extension is considered based on simulations performed with the PYTHIA 8 event generator in the context of $\rm pp$ collisions at $\sqrt{s}=13.6$ TeV but can be trivially applied to any other scenario and implemented in measurements of correlated particle production in $\rm pp$, pA, and AA collisions at colliders. Correlation of selected flavor balancing pairs are examined as function of the produced charged particle multiplicity. One finds that the amplitude of the correlations increases monotonically with the number of balanced flavors and the actual flavor content of correlated particles.
\end{abstract}

\maketitle

\section{Introduction}
\label{sec:introduction}

Differential correlation functions, and in particular unified balance functions (BFs) constitute a powerful  tool  to investigate the evolution of particle production and transport in heavy-ion collisions~\cite{Bass:2000az,Pratt:2002BFLH,Jeon:2002BFCF,Pruneau:2022brh,Pruneau:2023zhl,Pruneau:2024jpa}. 
They are nominally sensitive to the existence  of a long lived isentropic expanding quark--gluon plasma (QGP) as well as to the radial expansion dynamics of the matter formed in AA  collisions~\cite{Voloshin:2006TRE,Pruneau:2007ua,Bozek:2005BFTF,Pratt:2019pnd} and to the formation of clusters~\cite{Pratt:2018ebf}. 
The azimuthal width of hadron BFs, particularly heavier hadrons, is also sensitive to  the  diffusivity of light quarks~\cite{S.GavinAPHA:2006Diffusion,Jeon:2002BFCF,Pratt:2019pnd}.  
Furthermore, it has also been argued that mixed species balance functions, i.e., involving distinct species of hadrons, are sensitive to QGP susceptibilities near the phase transition~\cite{Pratt:2015jsa,Pruneau:2019BNC,Pruneau:2023cea}. 
General charge BFs  offer the distinctive advantage of featuring a sum rule: they integrate to unity in full acceptance calculations and measurements. Yet, their use may be somewhat restrictive because unlike-sign (unlike-strangeness, etc) correlation functions must be compared to like-sign (like-strangeness, etc) correlation functions. It is of interest, in particular,  to consider whether one can extend charge/strangeness/baryon BFs to more generic correlation functions exploiting   (approximate) quark flavor balancing to study correlated particle production dynamics in heavy-ion collisions and establish, potentially, the degree to which thermalization/equilibration is achieved in pp, pA, and AA collisions. 

In this work, we explore the notion of quark-flavor balancing, i.e., how correlation functions between particles featuring one, two, or three balancing quark flavors behave in relation to one  another. We specifically compare the strength and width of longitudinal correlations of hadrons containing    up ($\rm u$) and  anti-up ($\rm \bar u$),  down ($\rm d$) and  anti-down ($\rm \bar d$), or  strange ($\rm s$) and  anti-strange ($\rm \bar s$) quarks.
To this end, we examine correlations between hadron pairs featuring a single, two, or three balancing flavors.
Our study is based on pp collisions simulated with the Monash tune of the PYTHIA 8 event generator~\cite{Bierlich:2022pfr}
but shall eventually be extended to other event generators to determine the sensitivity of flavor balancing correlations to different quark production and recombination scenarios. 
We compute differential two-particle normalized cumulants and investigate  how their strength and shape evolve  with the flavor content and the number of balanced quark flavors.

While the ``upness" and ``downness" of quarks are not conserved, in general, due to charged currents in electroweak interactions, they are strictly conserved by the QCD Lagrangian. As such, given the relative weakness of the electroweak interaction, one  expects, to first approximation, that it plays a relatively minor role in correlations or balancing of flavors $\rm q\bar q$ in pp collisions or even on the time scale of the lifetime of central nucleus--nucleus (AA) collisions. One can thus expect that a comparative study of the strength and shape of  correlations of hadron species involving no, one, two, or three balanced flavors, should also enable some sensitivity to the  dynamics of pp and AA collisions. Indeed, just as  measurements of charge balance functions are useful to probe the production, transport, and evolution of particles in AA collisions ~\cite{Bass:2000az,Pratt:2002BFLH,Jeon:2002BFCF,Pruneau:2022brh,Pruneau:2023zhl,Pruneau:2024jpa,Voloshin:2006TRE,Pruneau:2007ua,Bozek:2005BFTF,Pratt:2019pnd,Pratt:2018ebf,S.GavinAPHA:2006Diffusion,Jeon:2002BFCF,Pratt:2019pnd}, measurements of the strength and shape of mixed species correlation functions involving varying numbers of balanced quark flavors should also provide some sensitivity to the production and transport of quark flavors as well as the hadronization processes. 

Within the QCD Lagrangian, $\rm u$, $\rm \bar u$, $\rm d$, $\rm \bar d$, etc, pairs are produced locally with a flavor ($f$) diagonal term of the form $\bar {\rm q}_f \gamma^{\mu}D_{\mu}{\rm q}_f$. The relative momenta of the created $\rm q\bar q$ pair is thus nominally only determined by the energy scale of the process and should, to first order, determine the longitudinal rapidity difference between the hadrons eventually produced. One may also wonder whether the mass of the produced $\rm q\bar q$ pair has an impact on the spatial-rapidity difference at which such a pair is sparked out of the vacuum. Additionally, one must note that quarks may scatter and diffuse from their original longitudinal rapidity distribution. In the context of the formation of QGP, this scattering manifests in part as a rapid longitudinal expansion of the dense fireball produced by colliding nuclei. The longitudinal ``distance" between  $\rm q$ and $\rm \bar q$ is then modified by the longitudinal flow. Evidently, the production time of a pair should also have an impact on their longitudinal separation 
at freeze-out\cite{Bass:2000az,Pratt:2002BFLH,Jeon:2002BFCF}. The longitudinal correlation width of flavor balanced hadrons should thus indeed be sensitive, as for charge balancing, to the production and transport of the flavors.

The number of balanced flavors might also impact the width and strength of the correlations. The presence of a  one-flavor balancing hadron pair, e.g., a proton ($\rm p\equiv uud$) and a positively charged pion ($\rm \pi^+\equiv u\bar d$), requires only one $\rm q\bar q$ pair be flavor balanced. Such a hadron pair is thus only minimally constrained by flavor conservation. The $\rm q\bar q$ pair  may be produced at any time or anywhere within the collision process and quarks needed to produce specific hadrons can be picked up combinatorially from ``co-moving" quarks. One can then naively expect such correlations to be longitudinally broad and weak because the balancing pair may be produced anywhere and many quarks can be randomly ``chosen" to make up the hadrons. The situation is modified however when two or three balancing flavors are entering into the correlated hadrons. In a two flavor balancing correlation, the two balancing pair partners must be close enough to one another to enter into the composition of the final hadrons. That implies they have to be produced relatively close together in both space and time to lead to form hadrons. 
At the same time, the combinatorial pickup of quarks to complete the makeup of the hadrons is as well reduced because two completions have to be carried on. This trend is expected to be even stronger when three flavors are balanced. In this case, three balancing pairs have to be produced close enough in the phase space for their components being part of the final hadrons. The probability of random combinations for completing the forming hadrons is even more reduced by the increase of completion requirements. The longitudinal correlation width is then naively expected to be reduced and the correlation strength significantly enhanced. However, one may also expect differences based on the flavor of the quarks. Strange quarks (and by extension heavier charm and bottom quarks) being significantly heavier than up and down quarks, their relative momenta (or longitudinal rapidity difference) are expected to be  smaller at any given energy scale. Their production is also expected to decrease in probability with the effective and local temperature of the system. Flavor balancing of up, down, and strange quarks might thus feature measurable differences. 

While the above scenario is admittedly rather schematic and likely simplistic, it is interesting to consider how measurements of mixed hadron flavor balancing correlation functions might contribute to the understanding of particle production and transport in $\rm pp$ and AA collisions. This evidently requires models that explicitly conserve both energy-momentum and flavor numbers. We thus begin our exploration of flavor balancing correlation functions with the Monash tune of PYTHIA 8 event generator. To put such studies in a specific context, the correlation functions are computed for $\rm pp$ collisions at $\sqrt{s}=13.6$ TeV. Correlations are computed using normalized differential two-particle cumulants $R_2(\Delta\mathrm{y}, \Delta\varphi)$~\cite{Adam:2017ucq,Pruneau:2022brh}  for several combinations of one, two, and three flavor balancing hadrons.

This paper is organized as follows. In Sec.~\ref{sec:definitions}, we introduce the differential correlation functions used in the analysis presented in this work and provide  a brief description of the PYTHIA 8 event generator and the simulations carried out to obtain the correlation functions of interest. Section~\ref{sec:results} presents the main results of this work, i.e., a detailed comparison of the correlation functions  obtained for several  pairs of hadron species involving one, two, and three balancing flavors, while its summary is presented in Sec.~\ref{sec:summary}. 

\section{Correlation Function Definitions and Analysis Details}
\label{sec:definitions}

This work is based on the differential normalized two-particle cumulant $R_{2}(\Delta\mathrm{y}, \Delta\varphi)$ defined according to
\begin{equation}
\label{eq:r2}
R_2^{\alpha\beta}(\Delta\mathrm{y},\Delta\varphi)= \frac{\rho_2^{\alpha\beta}(\Delta\mathrm{y},\Delta\varphi)}
{ \rho_1^{\alpha}\otimes\rho_1^{\beta}(\Delta\mathrm{y},\Delta\varphi)} - 1 
\end{equation}
with $\rho_1^{\alpha}(y,\varphi)$ and $\rho_2^{\alpha\beta}(\Delta\mathrm{y},\Delta\varphi)$ representing single- and two-particle densities, respectively. 
The  denominator $\rho_1^{\alpha}\otimes\rho_1^{\beta}$ is computed according to techniques documented in prior works \cite{Ravan:2013lwa,ALICE:2018jco}. The labels $\alpha$ and $\beta$ represent the hadron species of interest.

The study is performed on simulated $\rm pp$ collisions data generated with the PYTHIA 8.3 MC event generator. A total of $\rm 300M$ non-diffractive $\rm{pp}$ collisions were generated with the generator Monash tune~\cite{Bierlich:2022pfr}. However, in order to put this work in context, the analysis  was restricted to collisions that feature at least one charged particle produced in  forward/backward rapidity ranges corresponding to ALICE V0/T0 detectors acceptance, here considered to provide a minimum bias trigger. The analysis is carried out based on this minimum bias trigger as well as within three multiplicity classes determined from the charged particle multiplicity in  the range $-3.7 < |\eta| < -1.7$ and $2.8 < |\eta| < 5.1$ and corresponding to collisions amounting to  $0-10\%$ (largest multiplicity), $10-40\%$ and $40-80\%$ (smallest multiplicity) of the interaction cross section. 
Additionally, the weak decays of $\Lambda$, $\Sigma$, $\Xi$, $\Omega$, $\mathrm{K}_{\rm S}$ and $\mathrm{K}_{\rm L}$ are switched off to enable a direct analysis of correlation yields involving $\Lambda$  and to avoid feed down from weak decays that might complicate the interpretation of the results. Correlation involving  particles  of interest can then be studied with maximal efficiency and no ambiguity. 
Suppression of these weak decays also enable a focus on ``primary" particles, i.e., those particles emanating directly from the collisions and predominantly produced by strong QCD interactions. 
Charged hadrons $(\pi, \mathrm{K}, \mathrm{p})$ and $\Lambda$ are accepted in the rapidity range of $\pm 0.8$ within full azimuth. All the particles and their corresponding anti-particles are identified based on  their PDG codes~\cite{ParticleDataGroup:2020ssz}. Furthermore, they are selected in the $0.2 \le p_{\rm T} < 2.0$ range to capture the bulk of the particle production. Hadron and hadron-pair densities are determined based on event-ensemble averages and combined according to Eq.~(\ref{eq:r2}), for selected combinations of $\pi$, $\mathrm{K}$, $\mathrm{p}$, and $\Lambda$, to determine the $R_2^{\alpha\beta}(\Delta\mathrm{y},\Delta\varphi)$ correlation functions.


To avoid the intricacy of computing uncertainties based on conventional error propagation techniques, statistical uncertainties are computed based on the sub-sample analysis method. PYTHIA 8 generated events are sub-divided into ten sub samples, which are analyzed independently. The reported correlation functions correspond to the average along the ten sub-samples and the statistical uncertainties to the standard deviation of that average. As the PYTHIA 8 generated data sample is relatively large, the statistical uncertainties on the amplitude of $R_2^{\alpha\beta}(\Delta\mathrm{y},\Delta\varphi)$ are typically smaller than $0.1$\% of the highest amplitude.


\section{Results}
\label{sec:results}

\begin{figure}[ht]
\includegraphics[width=\linewidth,trim={1mm 6mm 12mm 5mm},clip]{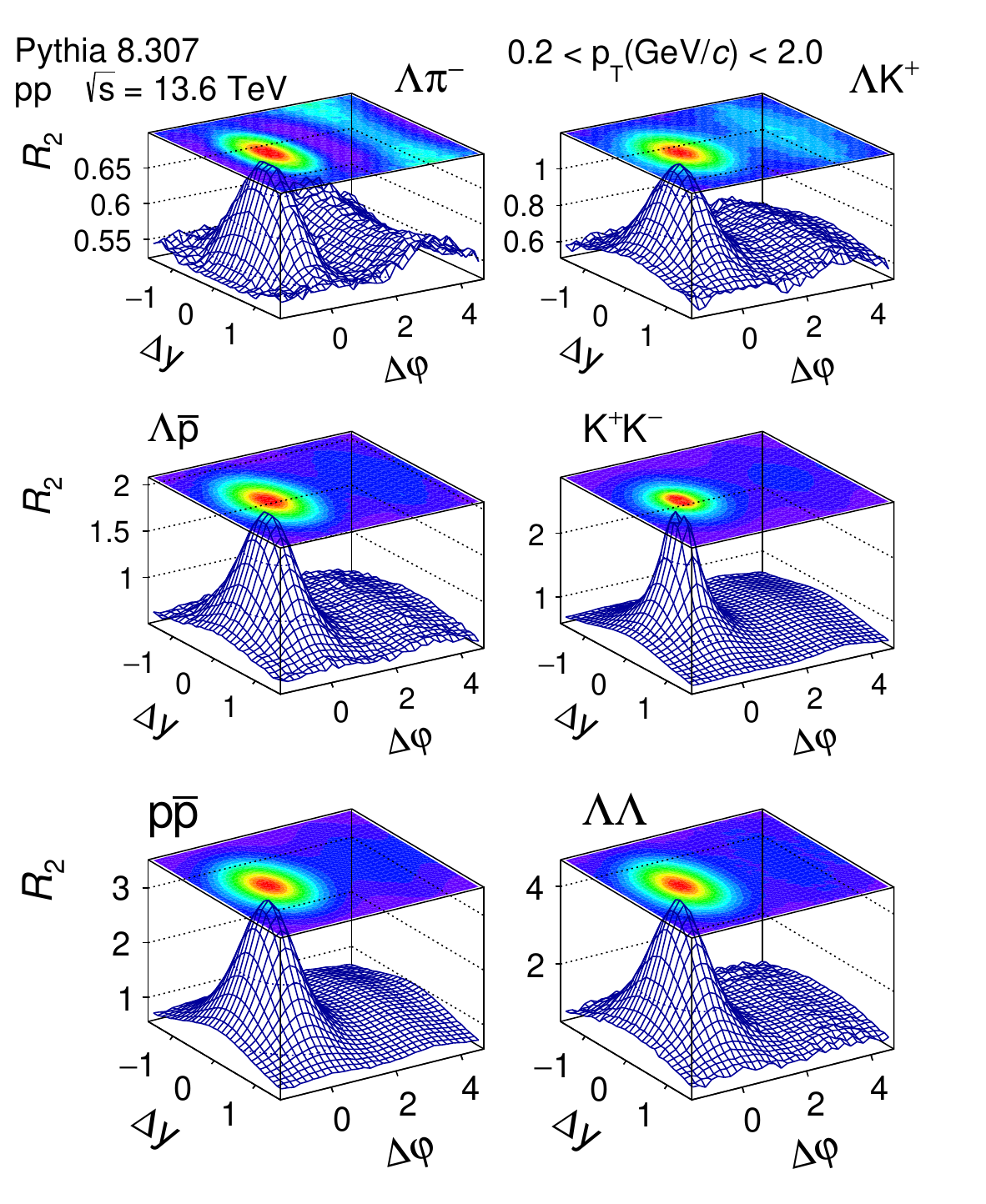}
\caption{Correlation functions $R_2(\Delta\mathrm{y},\Delta\varphi)$ of selected hadron pairs computed based on pp collisions at $\sqrt{s}=13.6$  TeV simulated with PYTHIA 8.
See text for details.}
\label{fig:can_3D_plots}
\end{figure}

Correlation functions $R_2$ of hadron pairs of interest are computed based on Eq.~(\ref{eq:r2}) as functions of $\Delta y$ and $\Delta \varphi$. Figure~\ref{fig:can_3D_plots} displays selected hadron-hadron $R_2(\Delta\mathrm{y},\Delta\varphi)$ correlation functions computed for minimum-bias PYTHIA 8 Monash tune pp collisions at $\sqrt{s}=13.6$ TeV. All correlation functions examined have similar basic characteristics. They feature a somewhat prominent near- side peak centered at $\Delta y=0$ and $\Delta \varphi=0$ as well as an elongated away-side ridge-like structure in rapidity difference, $\Delta y$, centered at $\Delta \varphi=\pi$. The near-side peaks of the correlation functions have maximum amplitudes and widths that depend on the hadron pair considered. The away-side ridge also features amplitudes and shapes that vary with hadron  pairs. Some of these are convex and quite strong, whereas others are concave and of relatively weak amplitude. 

Correlations of  $\Lambda\pi^{-}$ and $\Lambda\mathrm{K}^{+}$ shown in the top panels of Fig.~\ref{fig:can_3D_plots} exhibit a strong near-side peak as well as rather large away-side ridge. By contrast, $R_{2}$ correlation functions of $\rm \Lambda\bar p$, $\rm K^+K^-$, $\rm p\bar{p}$, and $\Lambda\bar \Lambda$, displayed in the middle and bottom row of  Fig.~\ref{fig:can_3D_plots}, feature relatively low amplitude  away-side ridges 
at $\Delta\varphi =\pi$. This indicates that PYTHIA 8 favors the production of correlated $\rm K^+K^-$ and baryon anti-baryons pairs at small relative rapidity and azimuthal angles. Additionally note that the pairs  $\Lambda\pi^{-}$ and $\Lambda\mathrm{K}^{+}$ each feature one balancing flavor. The former balances upness, whereas the latter balances strangeness, but neither balance charge nor baryon number, yet they appear correlated in the phase-space thus apparently revealing the $\rm q\bar{q}$ correlation pertaining to the QCD interactions. Their correlated production must thus involve at least one other particle to conserve charge  and baryon number. The  amplitude of these correlations are thus relatively weak owing most likely to the no restrictions on the pair creation within the phase-space and to the multiple combinations of particles that can accompany the observed pair and satisfy charge, strangeness, and baryon number conservation.  By contrast, correlations involving $\mathrm{K}^+ \mathrm{K}^-$, $\Lambda\bar{\mathrm{p}}$, $\rm p\bar{\mathrm{p}}$ and $\Lambda\bar\Lambda$ are clearly dominated by same-side emission, a prominent near-side peak, and feature a relative much weaker away-side contribution. This indicates that PYTHIA 8 favors flavor balancing of these hadron via same-side emission, proximity in phase-space, rather than back-to-back. 
It should also be noted that $\rm K^+ K^-$ and $\rm \Lambda\bar p$ pairs balance two flavors while $\rm p\bar p$ and $\rm \Lambda\bar\Lambda$  each balance three flavors. 
It is clear from these selected examples  that the $R_2$ correlator is extremely sensitive to events dominated by different particle production mechanism involving different number of balanced quark flavors.

\begin{figure}[ht]
\includegraphics[width=\linewidth,trim={1mm 0mm 10mm 2mm},clip]{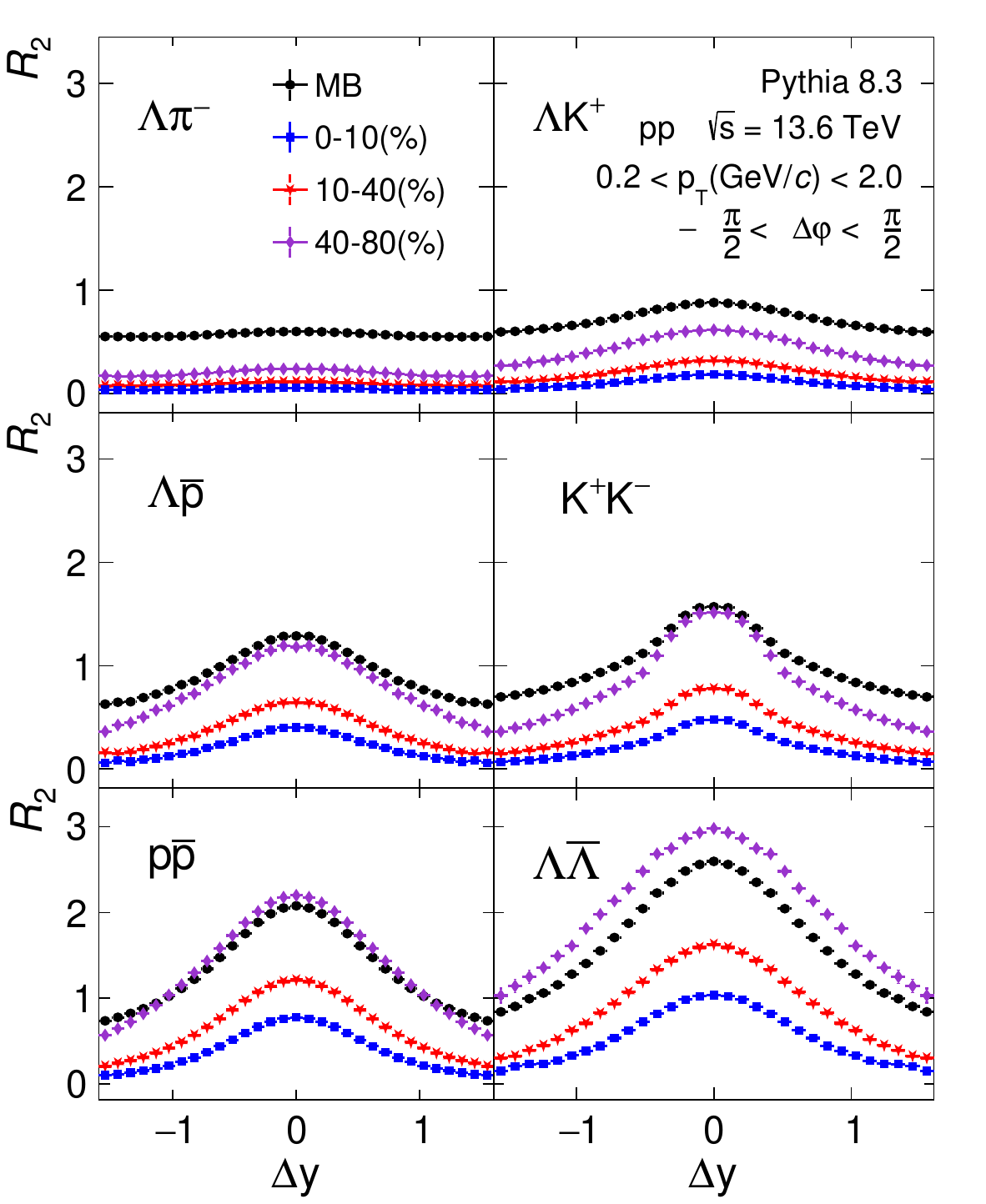}
\caption{Projections of the correlation function  $R_2(\Delta\mathrm{y},\Delta\varphi)$ on $\Delta y$ axis  for a selection of hadron pairs. Hadrons species are selected in the range $|y|<0.8$ and $0.2\le p_{\rm T}<2.0$ GeV/$c$. Projections are shown for minimum bias events (black points) and for three ranges of charged particle multiplicity emitted at forward and backward rapidities. See text for details.}
\label{fig:projX_plot}
\end{figure}

A more quantitative comparison of the $R_2(\Delta\mathrm{y},\Delta\varphi)$ of different hadron pairs is afforded with projections of the correlation functions onto the  $\Delta\mathrm{y}$ axis for $-\frac{\pi}{2} \leq \varphi \leq \frac{\pi}{2}$ shown in Fig.~\ref{fig:projX_plot}, which displays the projections of the correlation functions computed for minimum bias collisions as well as of those obtained from 0--10\%, 10--40\%, and 40--80\% multiplicity class collisions. One notes that all projections feature a more or less prominent peak centered at $\Delta\mathrm{y}=0$ with amplitudes, shapes, and widths that vary appreciably with the hadron pair considered. Also worth noting is that the peak amplitudes of the correlation functions decrease with increasing produced particle multiplicity since, in a high multiplicity environment, it is relatively easier to comply with the pair balancing creation and hadron completion requirements that can yield the hadrons of interest. Correlations obtained with 40-80\% have indeed the largest amplitude, followed by those from 10-40\% collisions, while the smallest amplitudes are observed for 0-10\% multiplicity class. Such a dependence on produced particle multiplicity is expected for correlation arising from collisions with multiple particle production sources.
If a single source is present, such as a cluster, a string, or a jet, produced particles are correlated by virtue of energy-momentum and quantum number conservation laws, including approximate flavor conservation. As the number of particle sources increases (e.g., several clusters, strings, jets, etc), the correlations are diluted by combinatorial pairs and the strength of $R_2$ functions accordingly decreases~\cite{Pruneau:2002yf}. One notes, indeed, that all hadron pairs show the same qualitative trend: their amplitudes decrease with an increase of  the produced multiplicity. 
One notes that the $R_{2}$ correlators in the $0-10\%$  multiplicity class feature slightly narrower near-side peaks  (centered at $\Delta\mathrm{y} = 0$ and $\Delta\varphi = 0$) as compared to those from the $40-80\%$ low multiplicity class. The strength, i.e., the amplitude $A_{R_{2}}$, of the correlator at $\Delta\mathrm{y} = 0$, is larger when strangeness balancing, $\mathrm{s}\bar{\mathrm{s}}$, is involved as compared to correlators involving only upness/downness flavor balancing. This is indicative of the fact that strange quark production is rarer as compared to u/d quarks: a gluon which splits into a $\mathrm{s}\bar{\mathrm{s}}$ pair requires at least $190 \ \mathrm{MeV}/\it{c}^{\mathrm{2}}$ which can only be achieved with momentum transfer of the same magnitude or larger, whereas much smaller momentum transfers can produce  $\mathrm{u}\bar{\mathrm{u}}$ or $\mathrm{d}\bar{\mathrm{d}}$ pairs. These comparison stands as a fruitful test of pQCD implemented in MC generators and should thus be verified based on detailed measurements in $\mathrm{p}\mathrm{p}$ collisions. 

\begin{figure}[ht]
\includegraphics[width=\linewidth,trim={0mm 0mm 0mm 0mm},clip]{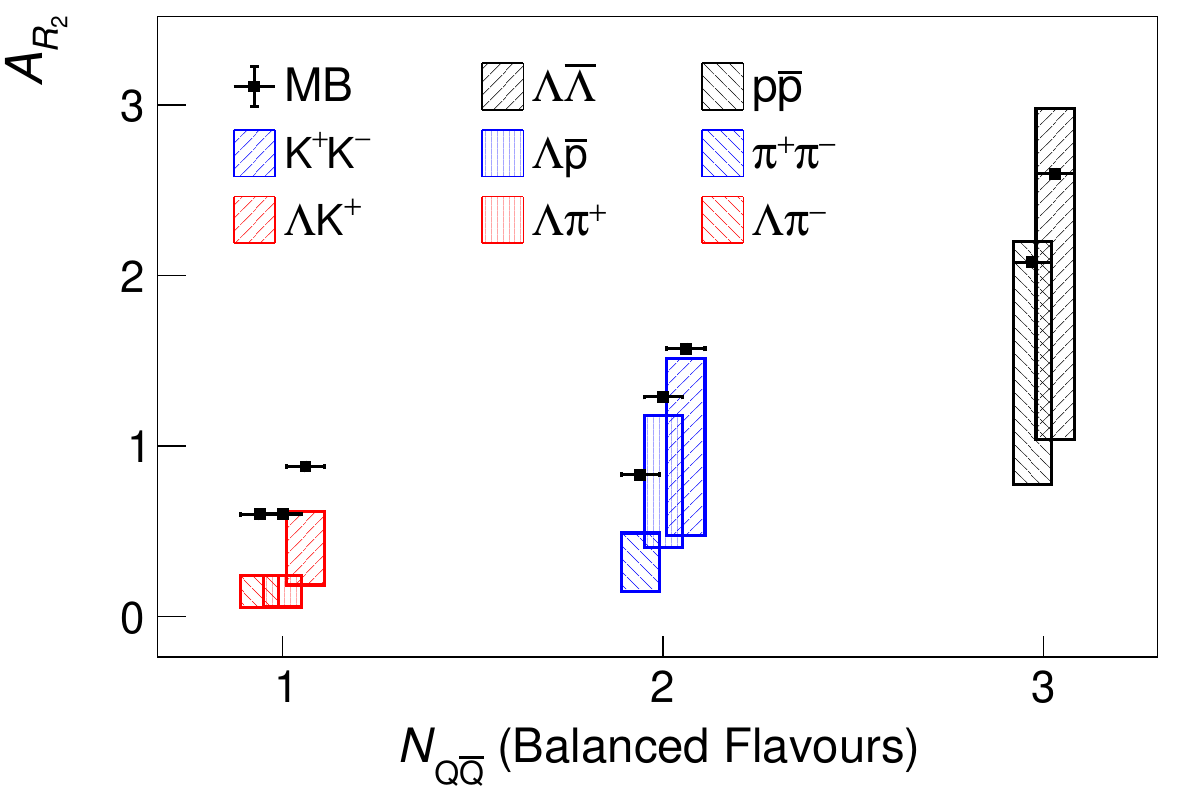}
\caption{Peak amplitude ($A_{R_{2}}$) of the $R_{2}$ correlation functions displayed in Fig.~\ref{fig:projX_plot} as a function of number of conserved and correlated quark-antiquark pairs $N_{\mathrm{Q}\bar{\mathrm{Q}}}$ in the correlated hadrons. The black square markers are the minimum bias points and the corresponding vertical bars show the evolution of $A_{R_{2}}$ through $0-10\%$ to $40-80\%$ multiplicity class as minima and maxima of the bar, respectively. (See text for details)}
\label{fig:amp_plot}
\end{figure}

We next examine the evolution of the amplitude of the correlation function projections displayed in Fig.~\ref{fig:projX_plot} and consider, in Fig.~\ref{fig:amp_plot}, their evolution with the number of balanced flavors and $\mathrm{p}\mathrm{p}$ multiplicity classes.
The peak amplitude of the correlation functions, denoted $A_{R_{2}}$,  corresponds  to the value of $R_{2}(\Delta\mathrm{y})$ at $\Delta\mathrm{y} = 0$. Black squares correspond to BFs obtained for minimum bias events, whereas colored bars represent the range of value obtained for  $0-10\%$ (bottom of range) and $40-80\%$  (top of range) multiplicity classes. In minimum-bias collisions, the  peak amplitude is large owing to the possibility of large fluctuations $\langle N(N-1)\rangle$ in the number $N$ of produced  particles  relative to the mean $\langle N\rangle$. The ratio $\langle N(N-1)\rangle/\langle N\rangle^2$, which determines the amplitude of the correlator $R_2$, is thus correspondingly large. However, when a small range of interaction cross section is selected, the fluctuations  are considerably reduced thereby yielding smaller values of $\langle N(N-1)\rangle/\langle N\rangle^2$. 
Additionally note that from higher to lower multiplicity class, the number of correlated pair of hadrons   decreases likely as a result of a reduction in the number of correlated particle sources.

Also note that the range of amplitudes (height of the bars) is somewhat  larger when strangeness is balanced, and there is a monotonic increase in the range of amplitudes with  the number of balanced flavours. This trend is also evident from the minimum bias markers and therefore the trend is preserved owing to the contribution of different particle production mechanism yielding from the quark-antiquark pairs originating from QCD interactions.

The observed trend vs. number of balanced flavors can be understood from a relatively simple, though schematic, arguments.  To first order, a balancing pair of $\mathrm{q}\bar{\mathrm{q}}$ might be expected to originate from a specific string but the quarks picked up to yield hadrons (i.e., color singlets) do not have to originate from the same string. There is thus, potentially, a rather large number of possibilities to produce hadrons with one balancing flavor. The correlation is thus diluted and weak. By contrast, as the number of balanced flavors is raised to two or three, the number of options is much reduced. Two balancing flavors may originate from the same string or neighboring strings, but the number of combinations is reduced. 
The probability of yielding these particular pairs is evidently reduced but the correlation is enhanced. There are only few ways to produce such two correlated hadrons. Random combinations of quarks are further reduced in the production of three balancing flavors. Whether the balancing flavor originate from the same string or neighboring strings is unclear at the outset but the $\mathrm{q}\bar{\mathrm{q}}$ cannot be chosen randomly. Each balanced flavor emerges from a specific pair creation process and the overall correlation strength is thus very strong even though the probability of the process itself is weak.

The above reasoning is evidently based in part on the notion of LUND strings used in  PYTHIA. Conceivably, other scenarios of particle production might lead to different behaviors or dependence of the correlation strength with the number of balancing quark flavors. On the other hand, if  LUND strings are in fact a reasonable mechanism of particle production in $\mathrm{p}\mathrm{p}$ collisions, one is left to wonder how such mechanism would be affected or modify the presence of a very  large number of close by or overlapping strings or in the context of the production of QGP.  Although the production  cross section of balancing pairs may be small, the expectation that these pairs are strongly correlated in $\mathrm{p}\mathrm{p}$ collisions but with potentially fast decreasing correlation strength in $\mathrm{A}\mathrm{A}$ collisions makes them an invaluable tool  to further understanding of particle production mechanisms in nuclear collisions.
Recent measurements of unlike-sign correlation functions in $\Delta \varphi$ by the ALICE Collaboration has shown some differences in strength and width of the correlation function on the near-side\cite{ALICE:2016jjg}.
The strength was observed to be highest for kaons, lower for protons and lambdas, and lowest for pions. The observed disagreement with the predictions of the PYTHIA 8 model, shown by the rapidity projections in this work, thus hints at the existence of additional mechanisms driving the particle production and its evolution. It also points to the excellent sensitivity of the two-particle correlators to the presence of such mechanisms.

\section{Summary}
\label{sec:summary}

We presented a study of flavor balancing correlation functions computed  for $\mathrm{p}\mathrm{p}$ collisions at $\sqrt{s}=13.7$ TeV simulated with the PYTHIA 8 model. The strength of the correlations, studied with the normalized differential two-particle cumulants $R_2$, is found to decrease monotonically with increasing charged particle multiplicity and to increase with the number of flavor balancing $\mathrm{q}\bar{\mathrm{q}}$ pairs. The correlation strength is also observed to depend on the quark flavor. Correlation function involving strange hadrons are  typically  stronger   than those featuring up and down quarks. The decrease with increase charged particle production is understood to be a relatively trivial dilution effect taking place when several distinct correlation source contribute to the particle production, whereas the dependence on the number of quark balancing flavor, though qualitatively understandable in the context of PYTHIA, may differ significantly in the context of other particle production models. One also evidently wonders how these correlations may evolve in the context of very large nucleus-nucleus collisions and the production of QGP. It shall then be of great interest to further  explore the behavior of flavor balancing correlations in the context of other models and more importantly in $\mathrm{p}\mathrm{p}$, $\mathrm{p}\mathrm{A}$, and $\mathrm{A}\mathrm{A}$ collisions at both RHIC and the LHC. It shall be interesting, in particular, to study how these correlation functions evolve with beam energy and system size. Also note in closing that while the correlations presented in this work focused on easily measurable mesons and baryons, advances in technologies and beam intensities, particularly at ALICE 3~\cite{ALICE:2803563} should also enable the study of charmness and bottomness balance functions.

\newenvironment{acknowledgement}{\relax}{\relax}
\begin{acknowledgement}
\section*{Acknowledgements}
This work was supported in part by the Department of Science and Technology (DST), Government of India, under  grants No. SR/MF/PS-02/2021-IITB (E-37126) and the United States Department of Energy, Office of Nuclear Physics (DOE NP), United States of America, under grant No. DE-FG02-92ER40713.
\end{acknowledgement}

\bibliography{main}

\providecommand{\noopsort}[1]{}\providecommand{\singleletter}[1]{#1}
\begin{thebibliography}{23}%
\makeatletter
\providecommand \@ifxundefined [1]{%
 \@ifx{#1\undefined}
}%
\providecommand \@ifnum [1]{%
 \ifnum #1\expandafter \@firstoftwo
 \else \expandafter \@secondoftwo
 \fi
}%
\providecommand \@ifx [1]{%
 \ifx #1\expandafter \@firstoftwo
 \else \expandafter \@secondoftwo
 \fi
}%
\providecommand \natexlab [1]{#1}%
\providecommand \enquote  [1]{``#1''}%
\providecommand \bibnamefont  [1]{#1}%
\providecommand \bibfnamefont [1]{#1}%
\providecommand \citenamefont [1]{#1}%
\providecommand \href@noop [0]{\@secondoftwo}%
\providecommand \href [0]{\begingroup \@sanitize@url \@href}%
\providecommand \@href[1]{\@@startlink{#1}\@@href}%
\providecommand \@@href[1]{\endgroup#1\@@endlink}%
\providecommand \@sanitize@url [0]{\catcode `\\12\catcode `\$12\catcode `\&12\catcode `\#12\catcode `\^12\catcode `\_12\catcode `\%12\relax}%
\providecommand \@@startlink[1]{}%
\providecommand \@@endlink[0]{}%
\providecommand \url  [0]{\begingroup\@sanitize@url \@url }%
\providecommand \@url [1]{\endgroup\@href {#1}{\urlprefix }}%
\providecommand \urlprefix  [0]{URL }%
\providecommand \Eprint [0]{\href }%
\providecommand \doibase [0]{http://dx.doi.org/}%
\providecommand \selectlanguage [0]{\@gobble}%
\providecommand \bibinfo  [0]{\@secondoftwo}%
\providecommand \bibfield  [0]{\@secondoftwo}%
\providecommand \translation [1]{[#1]}%
\providecommand \BibitemOpen [0]{}%
\providecommand \bibitemStop [0]{}%
\providecommand \bibitemNoStop [0]{.\EOS\space}%
\providecommand \EOS [0]{\spacefactor3000\relax}%
\providecommand \BibitemShut  [1]{\csname bibitem#1\endcsname}%
\let\auto@bib@innerbib\@empty
\bibitem [{\citenamefont {Bass}\ \emph {et~al.}(2000)\citenamefont {Bass}, \citenamefont {Danielewicz},\ and\ \citenamefont {Pratt}}]{Bass:2000az}%
  \BibitemOpen
  \bibfield  {author} {\bibinfo {author} {\bibfnamefont {S.~A.}\ \bibnamefont {Bass}}, \bibinfo {author} {\bibfnamefont {P.}~\bibnamefont {Danielewicz}}, \ and\ \bibinfo {author} {\bibfnamefont {S.}~\bibnamefont {Pratt}},\ }\href {\doibase 10.1103/PhysRevLett.85.2689} {\bibfield  {journal} {\bibinfo  {journal} {Phys.\ Rev.\ Lett.}\ }\textbf {\bibinfo {volume} {85}},\ \bibinfo {pages} {2689} (\bibinfo {year} {2000})}\BibitemShut {NoStop}%
\bibitem [{\citenamefont {Pratt}(2002)}]{Pratt:2002BFLH}%
  \BibitemOpen
  \bibfield  {author} {\bibinfo {author} {\bibfnamefont {S.}~\bibnamefont {Pratt}},\ }\href {\doibase 10.1016/S0375-9474(01)01421-X} {\bibfield  {journal} {\bibinfo  {journal} {Nucl. Phys. A}\ }\textbf {\bibinfo {volume} {698}},\ \bibinfo {pages} {531} (\bibinfo {year} {2002})}\BibitemShut {NoStop}%
\bibitem [{\citenamefont {Jeon}\ and\ \citenamefont {Pratt}(2002)}]{Jeon:2002BFCF}%
  \BibitemOpen
  \bibfield  {author} {\bibinfo {author} {\bibfnamefont {S.}~\bibnamefont {Jeon}}\ and\ \bibinfo {author} {\bibfnamefont {S.}~\bibnamefont {Pratt}},\ }\href {\doibase 10.1103/PhysRevC.65.044902} {\bibfield  {journal} {\bibinfo  {journal} {Phys. Rev. C}\ }\textbf {\bibinfo {volume} {65}},\ \bibinfo {pages} {044902} (\bibinfo {year} {2002})},\ \Eprint {http://arxiv.org/abs/hep-ph/0110043} {arXiv:hep-ph/0110043} \BibitemShut {NoStop}%
\bibitem [{\citenamefont {Pruneau}\ \emph {et~al.}(2023{\natexlab{a}})\citenamefont {Pruneau}, \citenamefont {Gonzalez}, \citenamefont {Hanley}, \citenamefont {Marin},\ and\ \citenamefont {Basu}}]{Pruneau:2022brh}%
  \BibitemOpen
  \bibfield  {author} {\bibinfo {author} {\bibfnamefont {C.}~\bibnamefont {Pruneau}}, \bibinfo {author} {\bibfnamefont {V.}~\bibnamefont {Gonzalez}}, \bibinfo {author} {\bibfnamefont {B.}~\bibnamefont {Hanley}}, \bibinfo {author} {\bibfnamefont {A.}~\bibnamefont {Marin}}, \ and\ \bibinfo {author} {\bibfnamefont {S.}~\bibnamefont {Basu}},\ }\href {\doibase 10.1103/PhysRevC.107.054915} {\bibfield  {journal} {\bibinfo  {journal} {Phys. Rev. C}\ }\textbf {\bibinfo {volume} {107}},\ \bibinfo {pages} {054915} (\bibinfo {year} {2023}{\natexlab{a}})},\ \Eprint {http://arxiv.org/abs/2211.10770} {arXiv:2211.10770 [hep-ph]} \BibitemShut {NoStop}%
\bibitem [{\citenamefont {Pruneau}\ \emph {et~al.}(2023{\natexlab{b}})\citenamefont {Pruneau}, \citenamefont {Gonzalez}, \citenamefont {Hanley}, \citenamefont {Marin},\ and\ \citenamefont {Basu}}]{Pruneau:2023zhl}%
  \BibitemOpen
  \bibfield  {author} {\bibinfo {author} {\bibfnamefont {C.}~\bibnamefont {Pruneau}}, \bibinfo {author} {\bibfnamefont {V.}~\bibnamefont {Gonzalez}}, \bibinfo {author} {\bibfnamefont {B.}~\bibnamefont {Hanley}}, \bibinfo {author} {\bibfnamefont {A.}~\bibnamefont {Marin}}, \ and\ \bibinfo {author} {\bibfnamefont {S.}~\bibnamefont {Basu}},\ }\href {\doibase 10.1103/PhysRevC.107.014902} {\bibfield  {journal} {\bibinfo  {journal} {Phys. Rev. C}\ }\textbf {\bibinfo {volume} {107}},\ \bibinfo {pages} {014902} (\bibinfo {year} {2023}{\natexlab{b}})}\BibitemShut {NoStop}%
\bibitem [{\citenamefont {Pruneau}\ \emph {et~al.}(2024{\natexlab{a}})\citenamefont {Pruneau}, \citenamefont {Basu}, \citenamefont {Gonzalez}, \citenamefont {Hanley}, \citenamefont {Marin}, \citenamefont {Dobrin},\ and\ \citenamefont {Manea}}]{Pruneau:2024jpa}%
  \BibitemOpen
  \bibfield  {author} {\bibinfo {author} {\bibfnamefont {C.}~\bibnamefont {Pruneau}}, \bibinfo {author} {\bibfnamefont {S.}~\bibnamefont {Basu}}, \bibinfo {author} {\bibfnamefont {V.}~\bibnamefont {Gonzalez}}, \bibinfo {author} {\bibfnamefont {B.}~\bibnamefont {Hanley}}, \bibinfo {author} {\bibfnamefont {A.}~\bibnamefont {Marin}}, \bibinfo {author} {\bibfnamefont {A.~F.}\ \bibnamefont {Dobrin}}, \ and\ \bibinfo {author} {\bibfnamefont {A.}~\bibnamefont {Manea}},\ }\href {\doibase 10.1103/PhysRevC.109.064913} {\bibfield  {journal} {\bibinfo  {journal} {Phys. Rev. C}\ }\textbf {\bibinfo {volume} {109}},\ \bibinfo {pages} {064913} (\bibinfo {year} {2024}{\natexlab{a}})},\ \Eprint {http://arxiv.org/abs/2403.13007} {arXiv:2403.13007 [hep-ph]} \BibitemShut {NoStop}%
\bibitem [{\citenamefont {Voloshin}(2006)}]{Voloshin:2006TRE}%
  \BibitemOpen
  \bibfield  {author} {\bibinfo {author} {\bibfnamefont {S.~A.}\ \bibnamefont {Voloshin}},\ }\href {\doibase 10.1016/j.physletb.2005.11.024} {\bibfield  {journal} {\bibinfo  {journal} {Phys. Lett. B}\ }\textbf {\bibinfo {volume} {632}},\ \bibinfo {pages} {490} (\bibinfo {year} {2006})},\ \Eprint {http://arxiv.org/abs/nucl-th/0312065} {arXiv:nucl-th/0312065} \BibitemShut {NoStop}%
\bibitem [{\citenamefont {Pruneau}\ \emph {et~al.}(2008)\citenamefont {Pruneau}, \citenamefont {Gavin},\ and\ \citenamefont {Voloshin}}]{Pruneau:2007ua}%
  \BibitemOpen
  \bibfield  {author} {\bibinfo {author} {\bibfnamefont {C.~A.}\ \bibnamefont {Pruneau}}, \bibinfo {author} {\bibfnamefont {S.}~\bibnamefont {Gavin}}, \ and\ \bibinfo {author} {\bibfnamefont {S.~A.}\ \bibnamefont {Voloshin}},\ }\href {\doibase 10.1016/j.nuclphysa.2008.01.031} {\bibfield  {journal} {\bibinfo  {journal} {Nucl. Phys. A}\ }\textbf {\bibinfo {volume} {802}},\ \bibinfo {pages} {107} (\bibinfo {year} {2008})},\ \Eprint {http://arxiv.org/abs/0711.1991} {arXiv:0711.1991 [nucl-ex]} \BibitemShut {NoStop}%
\bibitem [{\citenamefont {Bozek}(2005)}]{Bozek:2005BFTF}%
  \BibitemOpen
  \bibfield  {author} {\bibinfo {author} {\bibfnamefont {P.}~\bibnamefont {Bozek}},\ }\href {\doibase 10.1016/j.physletb.2005.01.072} {\bibfield  {journal} {\bibinfo  {journal} {Phys. Lett. B}\ }\textbf {\bibinfo {volume} {609}},\ \bibinfo {pages} {247} (\bibinfo {year} {2005})},\ \Eprint {http://arxiv.org/abs/nucl-th/0412076} {arXiv:nucl-th/0412076} \BibitemShut {NoStop}%
\bibitem [{\citenamefont {Pratt}\ and\ \citenamefont {Plumberg}(2020)}]{Pratt:2019pnd}%
  \BibitemOpen
  \bibfield  {author} {\bibinfo {author} {\bibfnamefont {S.}~\bibnamefont {Pratt}}\ and\ \bibinfo {author} {\bibfnamefont {C.}~\bibnamefont {Plumberg}},\ }\href {\doibase 10.1103/PhysRevC.102.044909} {\bibfield  {journal} {\bibinfo  {journal} {Phys. Rev. C}\ }\textbf {\bibinfo {volume} {102}},\ \bibinfo {pages} {044909} (\bibinfo {year} {2020})},\ \Eprint {http://arxiv.org/abs/1904.11459} {arXiv:1904.11459 [nucl-th]} \BibitemShut {NoStop}%
\bibitem [{\citenamefont {Pratt}\ and\ \citenamefont {Plumberg}(2019)}]{Pratt:2018ebf}%
  \BibitemOpen
  \bibfield  {author} {\bibinfo {author} {\bibfnamefont {S.}~\bibnamefont {Pratt}}\ and\ \bibinfo {author} {\bibfnamefont {C.}~\bibnamefont {Plumberg}},\ }\href {\doibase 10.1103/PhysRevC.99.044916} {\bibfield  {journal} {\bibinfo  {journal} {Phys. Rev. C}\ }\textbf {\bibinfo {volume} {99}},\ \bibinfo {pages} {044916} (\bibinfo {year} {2019})},\ \Eprint {http://arxiv.org/abs/1812.05649} {arXiv:1812.05649 [nucl-th]} \BibitemShut {NoStop}%
\bibitem [{\citenamefont {Abdel-Aziz}\ and\ \citenamefont {Gavin}(2006)}]{S.GavinAPHA:2006Diffusion}%
  \BibitemOpen
  \bibfield  {author} {\bibinfo {author} {\bibfnamefont {M.}~\bibnamefont {Abdel-Aziz}}\ and\ \bibinfo {author} {\bibfnamefont {S.}~\bibnamefont {Gavin}},\ }\href {\doibase 10.1556/APH.25.2006.2-4.43} {\bibfield  {journal} {\bibinfo  {journal} {Acta Phys. Hung.}\ }\textbf {\bibinfo {volume} {A25}},\ \bibinfo {pages} {515} (\bibinfo {year} {2006})}\BibitemShut {NoStop}%
\bibitem [{\citenamefont {Pratt}\ \emph {et~al.}(2015)\citenamefont {Pratt}, \citenamefont {McCormack},\ and\ \citenamefont {Ratti}}]{Pratt:2015jsa}%
  \BibitemOpen
  \bibfield  {author} {\bibinfo {author} {\bibfnamefont {S.}~\bibnamefont {Pratt}}, \bibinfo {author} {\bibfnamefont {W.~P.}\ \bibnamefont {McCormack}}, \ and\ \bibinfo {author} {\bibfnamefont {C.}~\bibnamefont {Ratti}},\ }\href {\doibase 10.1103/PhysRevC.92.064905} {\bibfield  {journal} {\bibinfo  {journal} {Phys.\ Rev.\ {\bf C}}\ }\textbf {\bibinfo {volume} {92}},\ \bibinfo {pages} {064905} (\bibinfo {year} {2015})}\BibitemShut {NoStop}%
\bibitem [{\citenamefont {Pruneau}(2019)}]{Pruneau:2019BNC}%
  \BibitemOpen
  \bibfield  {author} {\bibinfo {author} {\bibfnamefont {C.~A.}\ \bibnamefont {Pruneau}},\ }\href {\doibase 10.1103/PhysRevC.100.034905} {\bibfield  {journal} {\bibinfo  {journal} {Phys. Rev. C}\ }\textbf {\bibinfo {volume} {100}},\ \bibinfo {pages} {034905} (\bibinfo {year} {2019})},\ \Eprint {http://arxiv.org/abs/1903.04591} {arXiv:1903.04591 [nucl-th]} \BibitemShut {NoStop}%
\bibitem [{\citenamefont {Pruneau}\ \emph {et~al.}(2024{\natexlab{b}})\citenamefont {Pruneau}, \citenamefont {Gonzalez}, \citenamefont {Marin},\ and\ \citenamefont {Basu}}]{Pruneau:2023cea}%
  \BibitemOpen
  \bibfield  {author} {\bibinfo {author} {\bibfnamefont {C.}~\bibnamefont {Pruneau}}, \bibinfo {author} {\bibfnamefont {V.}~\bibnamefont {Gonzalez}}, \bibinfo {author} {\bibfnamefont {A.}~\bibnamefont {Marin}}, \ and\ \bibinfo {author} {\bibfnamefont {S.}~\bibnamefont {Basu}},\ }\href {\doibase 10.1103/PhysRevC.109.044904} {\bibfield  {journal} {\bibinfo  {journal} {Phys. Rev. C}\ }\textbf {\bibinfo {volume} {109}},\ \bibinfo {pages} {044904} (\bibinfo {year} {2024}{\natexlab{b}})},\ \Eprint {http://arxiv.org/abs/2310.07618} {arXiv:2310.07618 [hep-ex]} \BibitemShut {NoStop}%
\bibitem [{\citenamefont {Bierlich}\ \emph {et~al.}(2022)\citenamefont {Bierlich} \emph {et~al.}}]{Bierlich:2022pfr}%
  \BibitemOpen
  \bibfield  {author} {\bibinfo {author} {\bibfnamefont {C.}~\bibnamefont {Bierlich}} \emph {et~al.},\ }\href {\doibase 10.21468/SciPostPhysCodeb.8} {\bibfield  {journal} {\bibinfo  {journal} {SciPost Phys. Codeb.}\ }\textbf {\bibinfo {volume} {2022}},\ \bibinfo {pages} {8} (\bibinfo {year} {2022})},\ \Eprint {http://arxiv.org/abs/2203.11601} {arXiv:2203.11601 [hep-ph]} \BibitemShut {NoStop}%
\bibitem [{\citenamefont {Adam}\ \emph {et~al.}(2017{\natexlab{a}})\citenamefont {Adam} \emph {et~al.}}]{Adam:2017ucq}%
  \BibitemOpen
  \bibfield  {author} {\bibinfo {author} {\bibfnamefont {J.}~\bibnamefont {Adam}} \emph {et~al.} (\bibinfo {collaboration} {ALICE}),\ }\href {\doibase 10.1103/PhysRevLett.118.162302} {\bibfield  {journal} {\bibinfo  {journal} {Phys. Rev. Lett.}\ }\textbf {\bibinfo {volume} {118}},\ \bibinfo {pages} {162302} (\bibinfo {year} {2017}{\natexlab{a}})},\ \Eprint {http://arxiv.org/abs/1702.02665} {arXiv:1702.02665 [nucl-ex]} \BibitemShut {NoStop}%
\bibitem [{\citenamefont {Ravan}\ \emph {et~al.}(2014)\citenamefont {Ravan}, \citenamefont {Pujahari}, \citenamefont {Prasad},\ and\ \citenamefont {Pruneau}}]{Ravan:2013lwa}%
  \BibitemOpen
  \bibfield  {author} {\bibinfo {author} {\bibfnamefont {S.}~\bibnamefont {Ravan}}, \bibinfo {author} {\bibfnamefont {P.}~\bibnamefont {Pujahari}}, \bibinfo {author} {\bibfnamefont {S.}~\bibnamefont {Prasad}}, \ and\ \bibinfo {author} {\bibfnamefont {C.~A.}\ \bibnamefont {Pruneau}},\ }\href {\doibase 10.1103/PhysRevC.89.024906} {\bibfield  {journal} {\bibinfo  {journal} {Phys. Rev.}\ }\textbf {\bibinfo {volume} {C89}},\ \bibinfo {pages} {024906} (\bibinfo {year} {2014})},\ \Eprint {http://arxiv.org/abs/1311.3915} {arXiv:1311.3915 [nucl-ex]} \BibitemShut {NoStop}%
\bibitem [{\citenamefont {Acharya}\ \emph {et~al.}(2019)\citenamefont {Acharya} \emph {et~al.}}]{ALICE:2018jco}%
  \BibitemOpen
  \bibfield  {author} {\bibinfo {author} {\bibfnamefont {S.}~\bibnamefont {Acharya}} \emph {et~al.} (\bibinfo {collaboration} {ALICE}),\ }\href {\doibase 10.1103/PhysRevC.100.044903} {\bibfield  {journal} {\bibinfo  {journal} {Phys. Rev. C}\ }\textbf {\bibinfo {volume} {100}},\ \bibinfo {pages} {044903} (\bibinfo {year} {2019})},\ \Eprint {http://arxiv.org/abs/1805.04422} {arXiv:1805.04422 [nucl-ex]} \BibitemShut {NoStop}%
\bibitem [{\citenamefont {Zyla}\ \emph {et~al.}(2020)\citenamefont {Zyla} \emph {et~al.}}]{ParticleDataGroup:2020ssz}%
  \BibitemOpen
  \bibfield  {author} {\bibinfo {author} {\bibfnamefont {P.~A.}\ \bibnamefont {Zyla}} \emph {et~al.} (\bibinfo {collaboration} {Particle Data Group}),\ }\href {\doibase 10.1093/ptep/ptaa104} {\bibfield  {journal} {\bibinfo  {journal} {PTEP}\ }\textbf {\bibinfo {volume} {2020}},\ \bibinfo {pages} {083C01} (\bibinfo {year} {2020})}\BibitemShut {NoStop}%
\bibitem [{\citenamefont {Pruneau}\ \emph {et~al.}(2002)\citenamefont {Pruneau}, \citenamefont {Gavin},\ and\ \citenamefont {Voloshin}}]{Pruneau:2002yf}%
  \BibitemOpen
  \bibfield  {author} {\bibinfo {author} {\bibfnamefont {C.}~\bibnamefont {Pruneau}}, \bibinfo {author} {\bibfnamefont {S.}~\bibnamefont {Gavin}}, \ and\ \bibinfo {author} {\bibfnamefont {S.}~\bibnamefont {Voloshin}},\ }\href {\doibase 10.1103/PhysRevC.66.044904} {\bibfield  {journal} {\bibinfo  {journal} {Phys. Rev.}\ }\textbf {\bibinfo {volume} {C66}},\ \bibinfo {pages} {044904} (\bibinfo {year} {2002})},\ \Eprint {http://arxiv.org/abs/nucl-ex/0204011} {arXiv:nucl-ex/0204011 [nucl-ex]} \BibitemShut {NoStop}%
\bibitem [{\citenamefont {Adam}\ \emph {et~al.}(2017{\natexlab{b}})\citenamefont {Adam} \emph {et~al.}}]{ALICE:2016jjg}%
  \BibitemOpen
  \bibfield  {author} {\bibinfo {author} {\bibfnamefont {J.}~\bibnamefont {Adam}} \emph {et~al.} (\bibinfo {collaboration} {ALICE}),\ }\href {\doibase 10.1140/epjc/s10052-017-5129-6} {\bibfield  {journal} {\bibinfo  {journal} {Eur. Phys. J. C}\ }\textbf {\bibinfo {volume} {77}},\ \bibinfo {pages} {569} (\bibinfo {year} {2017}{\natexlab{b}})},\ \bibinfo {note} {[Erratum: Eur.Phys.J.C 79, 998 (2019)]},\ \Eprint {http://arxiv.org/abs/1612.08975} {arXiv:1612.08975 [nucl-ex]} \BibitemShut {NoStop}%
\bibitem [{\citenamefont {ALICE}(2022)}]{ALICE:2803563}%
  \BibitemOpen
  \bibfield  {author} {\bibinfo {author} {\bibfnamefont {C.}~\bibnamefont {ALICE}},\ }\href {https://cds.cern.ch/record/2803563} {\emph {\bibinfo {title} {{Letter of intent for ALICE 3: A next generation heavy-ion experiment at the LHC}}}},\ \bibinfo {type} {Tech. Rep.}\ (\bibinfo  {institution} {CERN},\ \bibinfo {address} {Geneva},\ \bibinfo {year} {2022})\ \bibinfo {note} {202 pages, 103 captioned figures, 19 tables},\ \Eprint {http://arxiv.org/abs/2211.02491} {arXiv:2211.02491} \BibitemShut {NoStop}%
\end{thebibliography}%
\end{document}